\documentclass[aps,pre,twocolumn,floatfix,showpacs,showkeys]{revtex4}
\usepackage{amsmath,amsfonts,amssymb,mathrsfs,graphicx}
\begin{document}
\title{Generalized Entropies and Statistical Mechanics}

\author{Fariel Shafee}
\affiliation{ Department of Physics\\ Princeton University\\
Princeton, NJ 08540\\ USA.} \email{fshafee@princeton.edu }

\begin {abstract}

We consider the problem of defining free energy and other
thermodynamic functions when the entropy is given as a general
function of the probability distribution, including that for
nonextensive forms. We find that the free energy, which is central
to the determination of all other quantities, can be obtained
uniquely numerically even when it is the root of a transcendental
equation. In particular we study the cases of the Tsallis form and a
new form proposed by us recently. We compare the free energy, the
internal energy and the specific heat of a simple system of two
energy states for each of these forms.

\end{abstract}

 \vspace*{1cm}

\maketitle

\section{Introduction}

We have recently \cite{FS1} proposed a new form of nonextensive
entropy which depends on a parameter similar to Tsallis entropy
\cite{TS1, TS2, PA1}, and in a similar limit approaches Shannon's
classical extensive entropy. We  have shown how the definition for
this new form of entropy can arise naturally in terms of mixing of
states in a phase cell when the cell is re-scaled, the parameter
being a measure of the rescaling, and how Shannon's coding theorem
\cite{NI1} elucidates such an approach.

In this paper we shall adopt a more general attitude and try to
develop the statistical mechanics of systems where the entropy is
defined almost arbitrarily. Such a 'designer' entropy \cite{LA} may
indeed be relevant in a specific context, but we shall not justify
here any specific form. The applicability of the Tsallis form which
leads to a Levy-type pdf found in many physical contexts is now well
established \cite{BE1, CO1, WO1} and in the earlier paper we have
commented about how our form may also be more relevant in a context
that demands a more stiff pdf. In this paper we concentrate on the
use of the pdf to obtain macroscopic quantities for general entropy
functions.

Central to the development of the statistical mechanics of a system
is the definition of the free energy, because it is related to the
normalization of the probability distribution function, which in
turn controls the behavior of all the macroscopic properties of the
ensemble. Hence we shall first establish a general prescription to
obtain the free energy, and then determine its value in a simple
physical case in terms of the temperature for Tsallis and the newer
form. We shall then use it to get the specific heat in both cases
and see how it changes with the change of the parameter at different
temperatures.

\section{ENTROPY AND PDF}

The pdf is found by optimizing the function

\begin{equation} \label{eq1}
L= S + \alpha ( 1 - \sum_i p_i) + \beta ( U - \sum_i p_i E_i)
\end{equation}

where $S$ is the entropy (whichever form), $U$ the internal energy
of the ensemble per constituent, $p_i$ is the probability of the
$i$-th state which has energy $E_i$, and $\alpha$ and $\beta$ are
the Lagrange multipliers associated with the normalization of the
pdf's $p_i$ and the conservation of energy.

Let us make the assumption that the entropy $S$ is a sum of the
components from all the states:

\begin{equation}\label{eq2}
S = \sum_i \phi(p_i)
\end{equation}

where $\phi$ is a generalized function which will in general be
different from the Shannon form:

\begin{eqnarray} \label{eq3}
S = - \langle \log(p)\rangle  \equiv - \sum_i p_i \log(p_i)
\end{eqnarray}

We get from optimizing $L$

\begin{equation} \label{eq4}
\phi'(p)= \alpha + \beta E_i
\end{equation}

with the simple solution

\begin{equation}  \label{eq5}
\phi(p_i) = (\alpha + \beta E_i) p_i
\end{equation}

the constant of integration vanishing because there can be no
contribution to the entropy from a state which has zero probability.

\section{FREE ENERGY}

We shall assume that the Helmholtz free energy $A$ is defined by

\begin{equation}\label{eq6}
\beta (U- A)= S
\end{equation}

Hence if $S$ is nonextensive, and $U$ is extensive, then $A$ must
also be nonextensive.

Using Eq.5 and Eq.6 get

\begin{equation}\label{eq7}
A = - \alpha/\beta
\end{equation}

Hence we get the relation for the pdf

\begin{equation} \label{eq8}
p_i = \psi^{-1} ( \beta(E_i- A))
\end{equation}

with the definition

\begin{equation}\label{eq9}
\psi(p) = \phi(p)/p
\end{equation}

Here we assume that the function $\psi$ can be inverted. This may
not always be the case for an arbitrary expression for the entropy,
at least not in a manageable closed form. Finally, $A$ can be
obtained from the constraint equation

\begin{eqnarray} \label{eq10}
\sum_i p_i = \sum_i \psi^{-1} (\beta (E_i -A)) = 1
\end{eqnarray}

Once $A$ has been determined, it may be placed in Eq.8 to get the
pdf $p_i$ for each of the states, all properly normalized, and we
can find $U$ and its derivative $C$, the specific heat. For the
simple system that we shall be concerned with, pressure and volume,
or their analogues, will not enter our considerations, and hence we
have only one specific heat, $C_v$, with $\beta$ now identified as
the inverse scale of energy, the temperature $T$,

\begin{equation}\label{eq11}
C= -\beta^2 \frac{\partial U}{ \partial \beta}
\end{equation}

\section{SHANNON AND TSALLIS THERMODYNAMICS}

The Shannon entropy Eq.3 immediately gives, using Eq.8 and Eq.9

\begin{eqnarray}\label{eq12}
p_i = e^{-\alpha - \beta E_i}= e^{\beta(A- E_i)}
\end{eqnarray}

so that Eq.10 yields the familiar expression for $A$

\begin{equation}\label{eq13}
A = - \log(Q)/\beta
\end{equation}

where $Q$ is the partition function

\begin{equation}\label{eq14}
Q = \sum_i e^{-\beta E_i}
\end{equation}

The exponential form of $p_i$ in Eq.12 allows the separation of the
$A$ dependent factor and hence such a simple expression for $A$ in
the case of Shannon entropy, giving us normal extensive
thermodynamics.

In the Tsallis case we have

\begin{eqnarray} \label{eq15}
S = -  \frac{\sum_i p_i^q -1}{q-1}= -\langle Log_q(p_i) \rangle
\end{eqnarray}

using the q-logarithm

\begin{equation} \label{eq16}
Log_q(p) = \frac{p^{q-1} -1}{q-1}
\end{equation}

and this is easily seen to lead to, using the symbol $\epsilon$ for
$(1-q)$

\begin{equation} \label{eq17}
p_i= \left(\frac{1}{1+ \epsilon \beta (E_i-A)}\right)^{1/\epsilon}
\end{equation}

We note that as it is no longer possible to separate out a common
$A$ dependent factor, we can no longer find an expression for $A$ in
terms of the partition function in the usual way. Instead it is
necessary to solve the normalization equation Eq.10. For a general
value of $\epsilon$ this will give an infinite number of roots, but
for values of $\epsilon$ corresponding to reciprocals of integers,
we shall have polynomial equations with a finite number of roots,
which too may be complex in general. We shall later see, at least
for the simple example we consider later, that a real and stable
root can be found that approaches the Shannon value of $A$ as
$\epsilon \rightarrow 0$, because in that limit the $Log_q(p)$
function in the definition of the Tsallis entropy also coincides
with the natural logarithm.

\section{THERMODYNAMICS OF THE NEW ENTROPY}

The entropy we proposed in ref.\cite{FS1} is given by

\begin{eqnarray}\label{eq18}
S = \frac{dM(q)}{dq} =- \sum_i p_i^q log(p_i)
\end{eqnarray}

with mixing probability \cite{FS1}

\begin{equation} \label{eq19}
M(q)= 1- \sum_i p_i^q
\end{equation}

We can also express it as the q-expectation value
\begin{eqnarray} \label{eq20}
S = - \langle log(p) \rangle_q \equiv \sum_i p_i^q  log(p_i)
\end{eqnarray}

Proceeding as in the case of Tsallis entropy we get, with $\epsilon=
1-q$,

\begin{equation}\label{eq21}
p_i = e^{-W(\epsilon \beta (E_i-A))/\epsilon}
\end{equation}

where $W$ is the Lambert function defined by

\begin{equation} \label{eq22}
W(z)e^{W(z)} = z
\end{equation}

Like the Tsallis case, here too we have to obtain $A$ by solving
numerically the transcendental Eq. 10. For the specific heat we get

\begin{equation} \label{eq23}
C = \beta^2 \sum_i  E_i (E_i-A)\frac{e^{-W_i( 1+1/\epsilon)}}{1+
W_i}
\end{equation}

where we have written for brevity $W_i = W \left(\epsilon \beta(E_i-
A)\right)$ and have used the following identities related to the
Lambert function \cite{CO}.

\begin{eqnarray}\label{eq2425}
W'(z) = \frac{W(z)}{z (1+ W(z))} \\
W(z)/z= e^{-W(z)}
\end{eqnarray}

As $W(z) \sim z$ for small $z$, we see that for small $\epsilon$ we
effectively get classical pdf and thermodynamics, as in the Tsallis
case. So the parameter $\epsilon$ is again a measure of the
departure from the standard statistical mechanics due to
nonextensivity. However, our nonextensivity is different
functionally from the Tsallis form \cite{FS1}, and the values of
$\epsilon$ in the two forms can only be compared in the limit of low
$\beta$. If we use the power series expansion of $W(z)$

\begin{equation}\label{eq26}
W(z) = \sum_{n=1} \frac{(-n)^{n-1} z^n}{n!}
\end{equation}

and compare that with the power series expansion for
$\log(1+z)$ in a form of the Tsallis $p_i$ similar to the new
entropy form

\begin{equation} \label{eq27}
p_i = e^{ -\log \left(1+ \epsilon_T \beta(E_i -
A)\right)/\epsilon_T}
\end{equation}

we get a cancelation of the parameter $\epsilon_n$ of the new
entropy and of Tsallis parameter $\epsilon_T$ in the first order, so
that both distributions approach the Shannon pdf, as we have already
mentioned, but if we demand equality in the second order, we get

\begin{equation} \label{eq28}
\epsilon_n= \frac{1}{2} \epsilon_T
\end{equation}

The third order difference between $W(Z)$ and $\log(1+z)$ is only
$\frac{1}{24}z^3$, and hence the difference between the Tsallis form
and our form of entropy will be detectable at rather low $T$,
i.e. high $\beta$.

\section{APPLICATION TO A SIMPLE SYSTEM}

Let us consider the simplest nontrivial system, where we have only
two energy eigenvalues $\pm E$, as in a spin-1/2 system. For a
noninteracting ensemble of systems we shall expect the standard
results \cite{PA} corresponding to Shannon entropy

\begin{eqnarray} \label{2932}
A = - \log [ 2 \cosh (\beta E)]/\beta \\
S = \log[2 \cosh(\beta E)]- \beta E \tanh (\beta E) \\
U = - E \tanh( \beta E)  \\
C = (\beta E)^2 /\cosh^2( \beta E)
\end{eqnarray}

\begin{figure}[ht!]
\includegraphics[width=8cm]{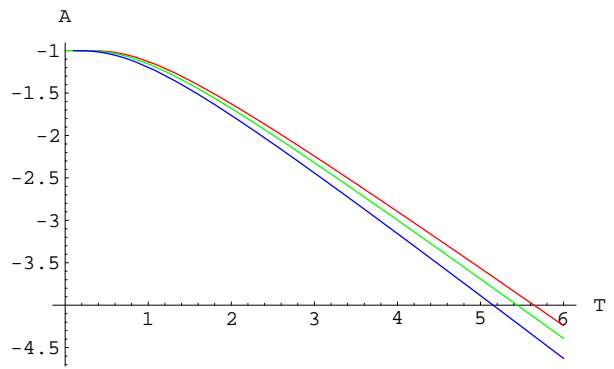}
\caption{\label{fig1} $A$ for Shannon (top), Tsallis with $\epsilon
=0.1$ (middle) and Tsallis with $\epsilon = 0.25$ (bottom)}
\end{figure}

\begin{figure}[ht!]
\includegraphics[width=8cm]{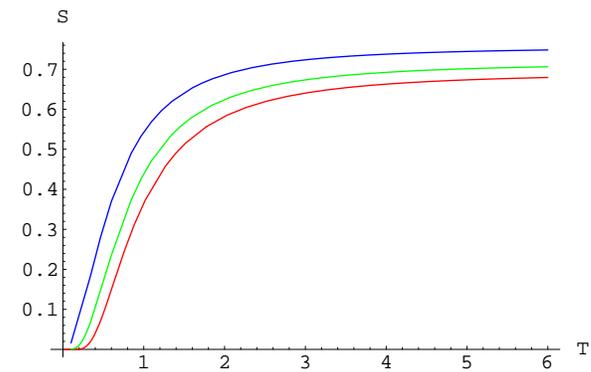}
\caption{\label{fig2} $S$ for same three entropy forms (from bottom
to top Shannon, Tsallis (0.1), Tsallis (0.25))}
\end{figure}

For the Tsallis entropy we shall take $\epsilon= 0.25, 0.10$ and
solve numerically. The values are shown in Figs 1-4. It is seen that
Tsallis entropy gives very similar shapes for all the variables and
for $\epsilon=0.10$ we get a fit much nearer to the Shannon form
than for $\epsilon= 0.25$.

The specific heat shows the typical Schottky form for a two-level
system.

\begin{figure}[ht!]
\includegraphics[width=8cm]{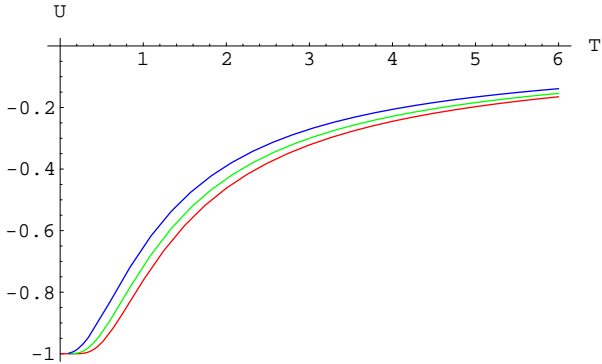}
\caption{\label{fig3} $U$ for same three entropy forms (same order
as for $S$)}
\end{figure}

\begin{figure}[ht!]
\includegraphics[width=8cm]{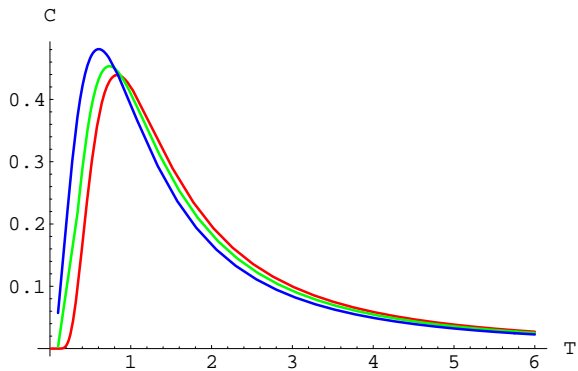}
\caption{\label{fig4} $C$ for same three entropy forms (peaks bottom
to top - Shannon, Tsallis (0.1), Tsallis (0.25))}
\end{figure}

Though the variables involve both $p_+$ and $p_-$, after finding $A$
we can replace $p_-$ by $1-p_+$ for faster execution of the
numerics.

For our new entropy too we first find the numerical value of $A$
from the normalization condition

\begin{equation} \label{eq33}
e^{-W_+/\epsilon}  + e^{-W_-/\epsilon}=1
\end{equation}

and then use this value to find $U$, $S$, and $C$. In Figs. 5-8 we
show the values corresponding to $\epsilon_n = 0.05$, which is half
the Tsallis parameter $\epsilon_T=0.10$ used, and also the Shannon
values. We note that only for the $S$ curve we have a perceptible
difference between Tsallis entropy and our new entropy at values of
$\beta$ near $1$.

\begin{figure}[ht!]
\includegraphics[width=8cm]{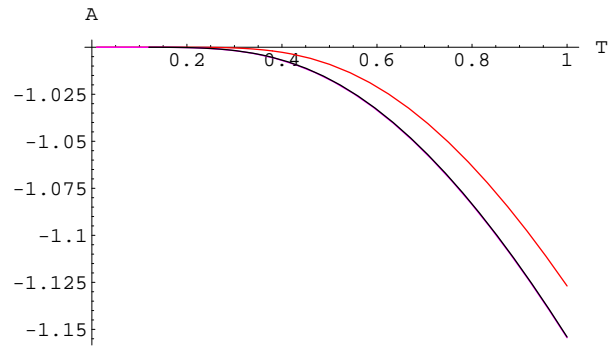}
\caption{\label{fig5} Comparison of $A$ for Shannon (apart), Tsallis
with $\epsilon=0.1$  and new entropy for $\epsilon=0.05$
(superposed)}
\end{figure}

\begin{figure}[ht!]
\includegraphics[width=8cm]{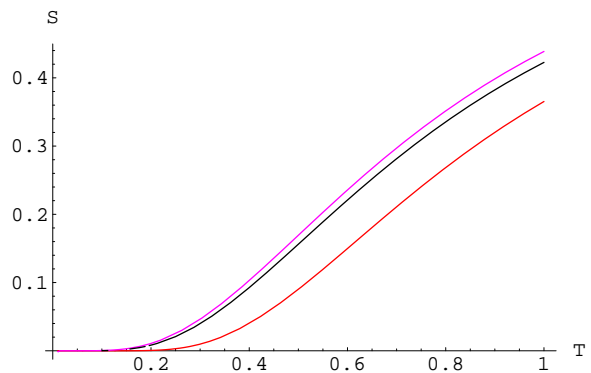}
\caption{\label{fig6} Comparison of $S$ for same three forms of
entropy (Shannon, Tsallis (0.1), new entropy(0.05) - Tsallis is
just over the new entropy ).}
\end{figure}

\begin{figure}[ht!]
\includegraphics[width=8cm]{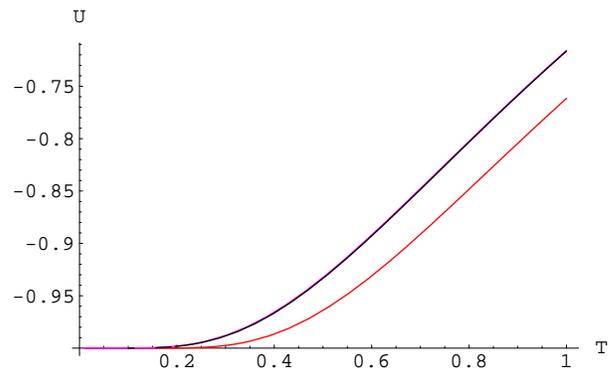}
\caption{\label{fig7} Comparison of $U$ for the same three entropies.
Shannon is separated, other two overlap.}
\end{figure}

\begin{figure}[ht!]
\includegraphics[width=8cm]{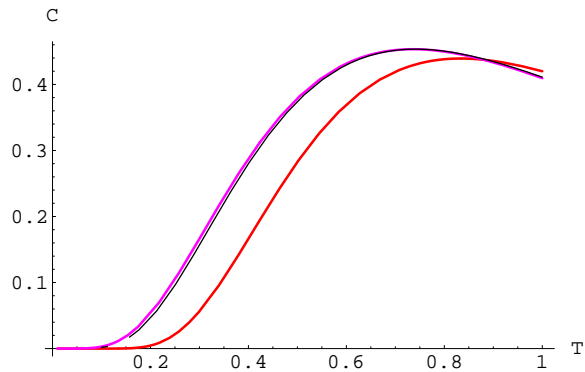}
\caption{\label{fig8}Specific heat C for same three entropies.
Again, only Shannon is separated.}
\end{figure}

\section{CONCLUSIONS}

We have presented above a simple prescription for finding the
important thermodynamic variables for any given form of the entropy
as a function of the pdf's of the states. We note that despite the
apparent complexity of the exponential or transcendental equations
determining the primary variable, the Helmholtz free energy $A$, it
is possible to numerically get stable values which approach the
expected Shannon values in the right limit of the parameter used,
both in the Tsallis case and in our case of the newly defined
entropy. For higher values of the parameter our entropy would give
values varying significantly from the Shannon or even the Tsallis
entropy, and there may be physical situations where that may indeed
be the desirable characteristic. But at low values the corresponding
values of the parameters for the two distributions produce virtually
completely overlapping graphs. The form of the entropy may be a
reflection of the effective interaction among the constituent
systems of the ensemble, the Shannon form being the limiting case of
the zero interaction case, and Tsallis or our form being results of
different forms of interaction with $\epsilon$ standing for a
coupling constant. It may be interesting to investigate more
complicated systems that are physically realizable and comparable.
It is noteworthy that most thermodynamic functions we have
considered here are not crucially dependent on the form of the
entropy with adjusted coupling, though the value of entropy itself
may vary significantly in the different formulations. This probably
indicates that the best way to discriminate between the suitability
of different definitions of entropy in different contexts may be in
comparing quantities that relate most directly to entropy.

\end{document}